\documentclass[preprint,prd,tightenlines,nofootinbib,floatfix]{revtex4}
\usepackage{graphicx}

\def \MSbar {\vbox{\hrule\kern 1pt\hbox{\rm MS}}}

\begin{document}

\begin{center}
{\Large   On the study of quark confinement and the relativistic flux tube model
}\\[1ex]
{Y. H. Yuan
\\ [1ex]
}

Department of Physics,
University of Wisconsin, 
Madison, WI 53706
\end{center}

\begin{abstract}

The scalar potential, time component vector potential and flux tube quark confinements are studied in this paper. We find that the predictions of scalar confinement and time component vector confinement are in considerable conflict with measured values while the flux-tube confinement works well to explain the experimental data. We also study the relativistic flux tube model. From the comparison of the exact numerical solution with the analytic approximation solution for heavy-light mesons, we find that the solutions are much more in agreement with each other for higher excited states since the deep radial limit is better satisfied.
 
\end{abstract}

\pacs{}
\maketitle

\section{Introduction}

The scalar potential quark confinement has been proposed as a significant model in hadron physics. In order to account for the observations of spin-orbit splitting, that the short range Coulomb potential was a Lorentz vector and the long range confining potential a Lorentz scalar was proposed by Henriques, Kellett, and Moorhouse \cite{Henriques}. There is no color magnetic field to act on the quarks' spin. As Buchm\"uller \cite{Buch} pointed out that a color electric flux tube should automatically yield the desired pure Thomas spin-orbit interaction because there is no color magnetic field in the quark rest frame. The Milan group clarified this question dramatically around 1990. Using the low velocity Wilson loop formalism pioneered by Eichten and Feinberg \cite{E&F}, and later by Gromes \cite{Gromes}, they found both the spin-dependent and spin-independent relativistic corrections in heavy onia\cite{Milan}. Their result was that the long-range spin-independent QCD corrections differed from those of scalar confinement, though the long-range spin-orbit corrections were the same pure Thomas ones given by scalar confinement. These results were subsequently verified by lattice simulations of QCD \cite{Bali}. The scalar quark confinement models remain popular in spite of the inconsistency with QCD, probably because of their relative ease of solution. We point out here that the predictions of scalar confinement also conflict directly with experiment. We study the scalar potential, time component vector potential and flux tube quark confinements and relativistic flux tube model in this paper. We investigate the dependence of heavy-light meson mass on the mass of the light quark. This paper is organized as follows. We exhibit a solution for scalar quark confinement in section II. In section III, we compare the predictions of the quark confinement with the experimental result and find that the predictions of the scalar confinement and the time component vector confinement are in considerable conflict with measured values while the flux-tube confinement works well to explain the experimental data. We review the relativistic flux tube model in section IV. Under two limit we solve the equations for the relativistic flux tube model in section V. An exact numerical solution is also demonstrated . In section VI, the comparison of the numerical solution with an analytical approximation solution is discussed. The conclusions are drawn in section VII. 

\section{Solution of the scalar quark confinement equation}
 
We consider a potential field consisting of central Lorentz scalar and time component vector fields, $S(r)$ and $V(r)$ respectively. To simplify our derivations, We assume that scalar confinement $S(r)=a|\mathbf{r}|=a r$ and a time component vector short range interaction, $V(\mathbf{r})=-{k\over |\mathbf{r}|}=-{k\over r}$. For a spinless quark $m$ moving in this potential field, we find

\begin{equation}
        H= \sqrt{{\bf p}^2+(m+ar)^2} - \frac{k}{r} .        \label{HamS}
\end{equation}
We wish to treat the short range parameter $k$ and the light quark mass $m$
as small perturbations.  In the case where both $k$ and $m$ vanish, the
zeroth order Hamiltonian is
\begin{equation}
                           H_0= \sqrt{{\bf p}^2+(ar)^2} .        \label{HamS0}
\end{equation}

where 

\begin{equation}
          {\bf p}^2={p_r}^2+{\ell(\ell +1)\over r^2}  .          \label{psph}
\end{equation}

The Hamiltonian $H_0$ has the same eigenstates as its square which, with the
replacement $p_r^2 \to -\,\frac{1}{r}\frac{d^2}{d
r^2 }r$, leads to the harmonic oscillator equation,
\begin{equation}
\frac{d^2u_0}{d r^2} + 
\left(E_0^2 - {\ell(\ell +1)\over r^2} - a^2r^2\right)u_0 = 0 ,
\label{HO}                         
\end{equation}
where in the limit of small $r$, $u_0\to r^{\ell+1}$  and $u_0$ is
normalized to
\begin{equation}
\int_0^{\infty} dr \, {\left| u_0 \right| }^2  =1 .    \label{intnorm}
\end{equation}
The solution for the wave function and eigenvalue is standard;
\begin{eqnarray}
u_0(r) &=& \mathcal{N}_{n,\ell}\, r^{\ell+1}\, e^{-\frac{1}{2}ar^2}\, 
L_{n-1}^{\ell+\frac{1}{2}}(ar^2)  ,      \label{soln0}\\
 \mathcal{N}_{n,\ell}^2 &=& \frac{2a^{\ell +\frac{3}{2}}\, (n-1)!}
{\Gamma(\ell+n+\frac{1}{2})} ,
\label{norm}\\
E_0^2 & = & 2a\,\left(\ell+2n-\frac{1}{2}\right) .          \label{eigen0}
\end{eqnarray}
Here $L_{n-1}^{\ell+\frac{1}{2}}(ar^2) $ is the usual Laguerre
polynomial, $n$ is a positive integer starting with $1$, and $\ell$ is a
non-negative integer starting with zero.

We now determine the effect of turning on $k$ and $m$ using the
Feynman-Hellmann theorem \cite{FH},
\begin{equation}
\partial E/\partial \lambda = \langle \partial H/\partial \lambda \rangle  ,
\label{FH}
\end{equation}
where $\lambda$ is any parameter of the Hamiltonian.  Taking the
expectation values using the $k=m=0$ wavefunctions (\ref{soln0}) will yield
an expansion in these parameters.  To leading order we have
\begin{equation}
 E=E_0-k \, \left\langle {r^{-1}}\right\rangle +\frac{ma}{E_0}\,
\langle r \rangle +\dots  \ .    \label{expan}
\end{equation}

The expectation values are worked out in general in the Appendix.  Here we
only consider the $m$ and $k$ dependence of the 1\textit{S} and 1\textit{P}
states (i.e., $n=1$ and $\ell=0$ and $1$).  The results are,
\begin{eqnarray}
\langle {r^{-1}} \rangle\strut_{1S} 
& = &2 \sqrt{\frac{a}{\pi} } , \label{1S/r}\\
\langle {r^{-1}} \rangle\strut_{1P} 
& = &\frac{4}{3} \sqrt{\frac{a}{\pi} }, \label{1P/r}\\
\langle r \rangle\strut_{1S} & = & \frac{2}{\sqrt{\pi a}}  , \label{1Sr}\\
\langle r \rangle\strut_{1P} & = & \frac{8}{3\sqrt{\pi a}} . \label{1Pr}
\end{eqnarray}

Using Eq.~(\ref{expan}), we obtain the $m$ dependence near $k=m=0$ ($m$
slope) for the energies,
\begin{eqnarray}
\frac{d\,E_{1S}}{dm} & = &  \frac{2}{\sqrt{3\pi}} =0.6515 ,  \label{mE1S} \\
\frac{d( E_{1P}-E_{1S})}{dm} 
& = &\frac{2} {\sqrt{3\pi}}\left(\frac{4}{\sqrt15}-1\right)= 0.0214 .
\label{mdiff}
\end{eqnarray}  
Again using Eq.~(\ref{expan}) we obtain the $k$ dependence near $k=m=0$
($k$ slope) for the energies,
\begin{eqnarray}
 \frac{1}{\sqrt a } \frac{d\,E_{1S}}{dk} 
& = & -2\sqrt{\frac{1}{\pi} } =~-1.128 ,  \label{kE1S} \\
 \frac{1}{\sqrt a } \frac{d( E_{1P}-E_{1S})}{dk} 
& = &\frac{2}{3}\sqrt{\frac{1}{\pi} } = 0.376 .  \label{kdiff}
\end{eqnarray}  

Next we compare our analytic values for
the slopes at $m=k=0$ in Eqs.~(\ref{mE1S}) to (\ref{kdiff}) with our
numerical method.  This step is important because a general analytic
solution of the spinless Salpeter equation is not known.  Only for the
remarkable case of a massless particle and linear scalar confinement, which
is equivalent to the non-relativistic harmonic oscillator, can one obtain
an analytic solution.  In the general case, one must rely on exact
numerical solutions. Some time ago, a
variational method, the Galerkin method, was introduced into particle physics to solve the
spinless Salpeter equation with a time component vector interaction\cite{Galerkin}.  This
very robust method is applicable to a wide range of differential and
integral equations.  The method has been sharpened over the years by many
authors \cite{others}.  One can cope with eigenvalue equations for
operators that are complicated functions of both momenta and coordinates,
such as the scalar confinement Hamiltonian (\ref{HamS}), by using basis
functions that can be Fourier transformed.  We have performed a careful
numerical solution of the eigenvalue equation for the Hamiltonian
(\ref{HamS}), for small $k$ and $m$, and found the $m$ and $k$ slopes. The
results are in excellent agreement with the values obtained by analytic
calculation in Eqs.~(\ref{mE1S}) through (\ref{kdiff}).

\section{Comparison of the predictions of different quark confinements with experimental results} \label{sec:experiment}

\subsection{Spin splitting is independent of light quark mass}
\label{subsec:spinsplit} 

We first use experimental data to demonstrate, rather conclusively, that
spin splittings within a given orbital angular momentum multiplet do not
depend on the light quark mass.  In particular, we consider several
heavy-light meson spin multiplets in which both the strange and non-strange
members have been observed \cite{PDG}.  First we examine the $D$ and $D_s$
type mesons with $\ell =0$ (\textit{S}-waves).  The hyperfine splittings
for $D_s$ and $D$ mesons are
\begin{eqnarray}
        D_s^* - D_s & = & 143.8~\pm 0.4~\textrm{MeV},  \nonumber     \\
        D_\pm^* - D_\pm & = & 140.64~\pm~0.10~\textrm{MeV}, \label{Dhyp} \\
        D_0^* - D_0 & = &  142.12~\pm 0.07~\textrm{MeV}. \nonumber
\end{eqnarray}
The corresponding \textit{S}-wave hyperfine splittings for $B$ type states are
\begin{eqnarray}
        B_s^* - B_s & = & 47.0~\pm 2.6 ~\textrm{MeV}, \nonumber \\
           B^* - B  & = &  45.78~\pm 0.35~\textrm{MeV}.        \label{Bhyp}
\end{eqnarray}
It is clear that the substitution of a strange quark for a non-strange
light quark changes the \textit{S}-wave hyperfine differences by at most a few MeV.

We next consider some measured \textit{P}-wave heavy-light spin splittings.  From
\cite{PDG} we find,
\begin{eqnarray}
      D_{s2} - D_{s1}    & = & 37.0~\pm 1.6~\textrm{MeV} , \nonumber \\
      D_2^0 - D_1^0      & = & 36.7~\pm 2.7~\textrm{MeV} ,   \label{DP} \\
      D_2^\pm  - D_1^\pm & = & 32~\pm 6 ~\textrm{MeV} . \nonumber
\end{eqnarray}

Again we note the apparent vanishing of light quark mass dependence this
time in a \textit{P}-wave spin splitting.  We conclude that spin splittings are only
weakly dependent on the light quark mass.
\subsection{The light quark mass dependence of a 1$P$-1$S$ difference} 
\label{subsec:massdiff}

In the preceding subsection we observed from experiment that both \textit{S}-wave and
\textit{P}-wave heavy-light spin splittings (within a spin multiplet) were
independent of the light quark mass.  We next consider the mass splittings
between pairs of states
corresponding to different orbital angular momenta and examine the light
quark mass dependence of this difference.  We choose the $D_1$ \textit{P}-wave state
and the pseudoscalar $D$ meson.  The best measurements are
\begin{equation}
  \Delta_u \ = \   D_1^0 - D^0 \ = \  557.5~\pm 2 ~ \textrm{MeV}.\label{SPD}
\end{equation}
When  the $u$ light quark is replaced by a strange quark, the corresponding
difference becomes
\begin{equation}
  \Delta_s \ = \ D_{s1} - D_s \ = \ 567.3~\pm 0.4~ \textrm{MeV}.\label{SPSD}
\end{equation}
The differences $\Delta_u$ and $\Delta_s$ are amazingly similar.  We see that
they differ by
\begin{equation}
 \Delta \ = \ \Delta_s -  \Delta_u \ = \ 9.8~\pm 2~\textrm{MeV}. \label{Diff}
\end{equation}

We demonstrated in Sec.~\ref{subsec:spinsplit} that both $S$- and $P$-wave spin splittings are, within error and isospin uncertainty,
independent of light quark mass.  We may therefore conclude that the
$m$-dependence of the difference $ \Delta_s- \Delta_u$ also represents the
$m$-dependence of the spin-averaged 1$P$-1$S$ excitation
energies.  To show this explicitly, we write the HL meson mass as the sum
of the heavy quark mass and the excitation energy,
\begin{equation}
M=m_Q +E .
\end{equation}
We then separate the excitation energy into spin-averaged and
spin-dependent parts,
\begin{equation}
E=E^{SA} +E^{SD} .
\end{equation}
As in Eqs.~(\ref{SPD}) and (\ref{SPSD}), we define $\Delta_s$ and
$\Delta_u$ to be the differences between $P$-wave ($J^P = 1^+$) and $S$-wave ($J^P = 0^-$) states, for strange and non-strange light quarks
respectively.  Using our observation that the spin-dependent parts are
essentially independent of light quark mass, we see that the spin dependent
parts will cancel in the difference $\Delta ={\Delta_s} -{\Delta_u}$. Thus
\begin{equation}
\Delta = (E_{1P} -E_{1S})_s -  (E_{1P} -E_{1S})_u 
\end{equation}
measures the light quark mass dependence of the spin-independent 1$P$-1$S$
level difference.

Two main results emerge from this subsection.  First, the light quark mass
dependence is only about 2\% of the measured level difference. This is a
useful tool in the analysis of heavy-light spectroscopy \cite{universal}.
Second, although it is small, $\Delta$ is definitely not zero.  We will use
the actual value (\ref{Diff}) to test the predictions of various
assumptions about the nature of quark confinement.
                             
\subsection{Comparison with experimental data} 
\label{sec:comparison}

We are now prepared to compare carefully the different confinement
predictions with experiment.  As we noted, the
$m$ dependences of all the heavy-light states are amazingly similar.  We
have also noted that this universal $m$
dependence is nearly satisfied in an analytical calculation.  In the
example with scalar confinement, the $m$ slope for the difference 1\textit{P}$-$1\textit{S} is
about 30 times smaller than each separate slope.  Furthermore, we noted in
Sec.~\ref{subsec:massdiff} that when one compares two heavy-light states, one
\textit{P}-wave and one \textit{S}-wave, the difference changes by less
than 2\% when a
non-strange light quark is replaced by a strange one.  This
change is not zero however, but for the $D_1$ and $D$ states has the
experimental value (\ref{Diff})
\begin{equation}
  \Delta \ = \ \Delta_s - \Delta_u \ = \ 9.8~\pm 2~\textrm{MeV} .
\end{equation}
In FIG.1 we show the numerical mass splittings of heavy-light
mesons for three different confinement scenarios, all with the same short
range energy $-k/r$, and the same $k=0.5$, as a function of light quark
mass. Each of the three confinement scenarios has the same asymptotic
confinement force, $a=0.18$~GeV$^2$.  The upper curve assumes linear scalar
confinement, the middle curve is the prediction of the relativistic flux
tube and the lower curve linear time component vector confinement. In the
scalar and  time component vector potentials, the potentials $S(r)$
and the long-range part of $V(r)$ respectively are $ar$.  In the case
relativistic flux tube model the string tension is $a$.
 
\begin{figure}[ht]
\includegraphics[width=\columnwidth,angle=0]{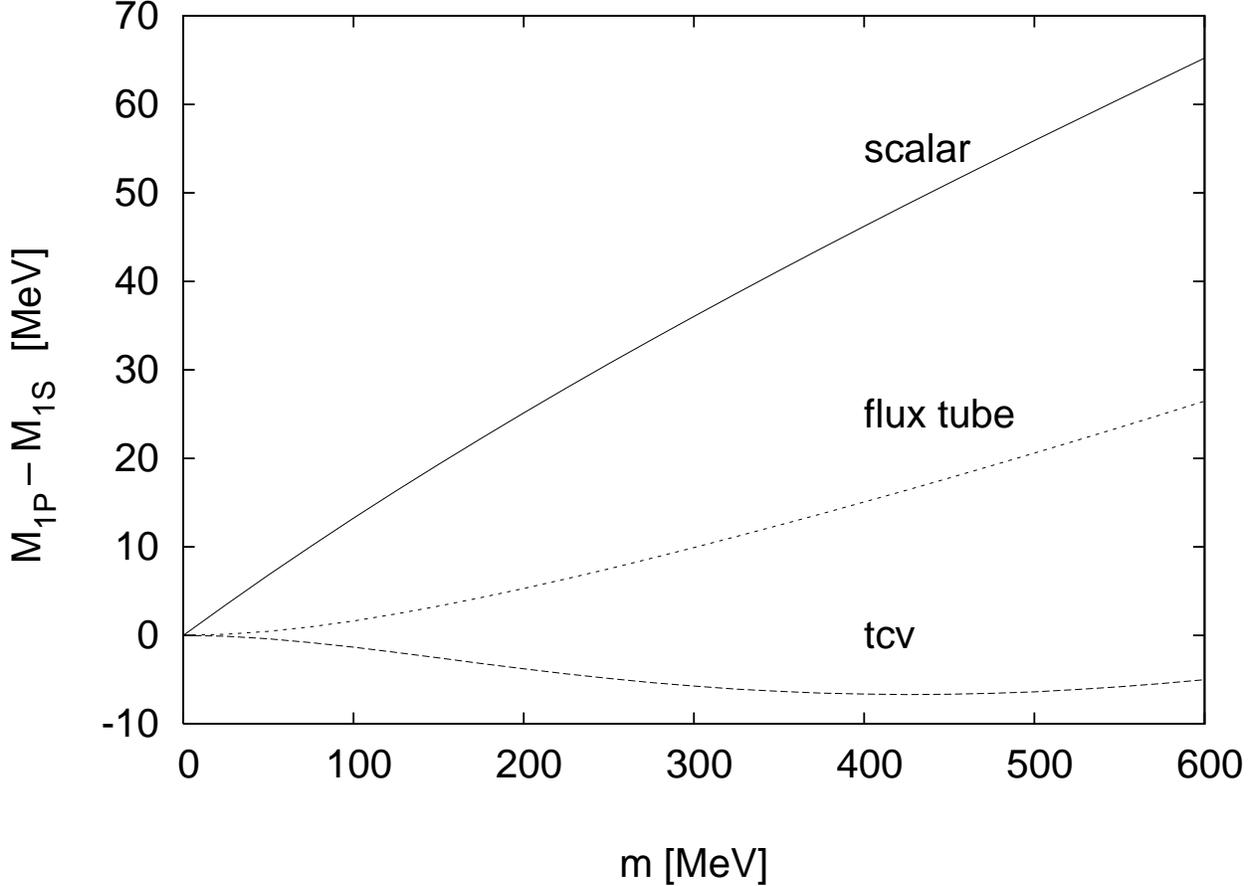}
\caption{Dependence of the 1\textit{P}$-$1\textit{S} energy level difference on the light quark
mass $m$. The numerical results compare different confinement mechanisms
(all with the same long range confinement force $a=0.18$~GeV$^2$ and all
three calculations assume the same short range Coulombic constant
$k=0.5$.)\label{fig:1}}
\end{figure}

We note that the scalar confinement has the most rapid increase of the
1\textit{P}$-$1\textit{S} difference as the light quark mass increases. The
flux tube confinement is intermediate and time component vector linear
confinement actually decreases slightly.  Using the reasonable values for
the strange quark mass (500 MeV) and the non-strange mass (300 MeV), we find the following values for $\Delta$,
\begin{eqnarray}
     \Delta_{\rm scalar}\ =\ 19~\textrm{MeV} ,   \\
     \Delta_{\rm flux~tube}\ =\ 10~\textrm{MeV} ,\\
     \Delta_{\rm TCV}\ =\ -1~\textrm{MeV}  .    
\end{eqnarray}
In comparing these values  to the experimental value given in Eq.~(\ref{Diff}),
\begin{equation}
     \Delta_{\rm exp}=~9.8~\pm 2~\textrm{MeV} ,
\end{equation}
we observe that both scalar confinement and time component vector
confinement are quite inconsistent with the experimental result.  However,
the flux tube confinement prediction is in agreement with the experimental
result. 

\section{Review of relativistic flux tube equations}

From the above study on the different quark confinement models, we observe that flux tube confinement prediction is in agreement with the experimental result. Next we continue our research on the flux tube model and concentrate on the relativistic flux tube (RFT) model\cite{lm}. The RFT model or QCD string play an important role in describing the relativistic meson states. Its fundamental assumption is that the QCD dynamical ground state for large quark separation consists of a rigid straight tube-like color flux structure connecting the quarks and for massless quarks it reduces to Nambu string.\cite{na} This model is consistent with both spin-dependent and spin-independent QCD expectations while the scalar confinement potential model\cite{bu} is not consistent with spin-independent QCD relativistic correction. Even though with this simple physical picture the RFT model coincides with the QCD relativistic correction derived from the rigorous Wilson loop formalism. We now review the spinless quarks and straight QCD string model and give the RFT equations. Consider a heavy-light (HL) meson made up of two spinless quarks, one heavy quark with mass M and one light quark with mass $ m_l$. They satisfy the conditions as $M\gg m_l\rightarrow 0$  and $ M\gg ar  $  where a is the confinement constant and r is the interquark separation. Since the center of mass remains close to the heavy quark in the HL meson, the constraint that the total momenta perpendidicular to the rotation axis and to the interquark axis vanish is automatically satisfied. For a heavy-light meson (HL), we obtain the equations
of QCD string as\cite{lm},

\begin{eqnarray} 
\frac Lr=W_r \gamma_\bot v_\bot +2arf(v_\bot)
 \label{e}\\ 
E=W_r \gamma_\bot +arS(v_\bot)+V(r)
\label{q} 
\end{eqnarray}

where$ L$ and $E $ are angular momentum and energy of the meson respectively, V(r) represents the short range potential. We also define the functions:

\begin{eqnarray} 
S(v_\bot) = \frac{arcsin \,v_\bot}{v_\bot}
\label{eq.3}\\
f(v_\bot)={1\over 4 v_\bot} [\,S(v_\bot) -\sqrt{1-v_\bot ^2}\,]
 \label{eq.4}\\
W_r =\sqrt{p_r ^2 +m^2}
\label{eq.5}\\
\gamma_\bot^2 = \frac{1}{1- v_\bot^2}
\label{eq.6}
\end{eqnarray}

\section{Solutions under two limits}

\vskip 0.3cm
\subsection{Circular Motion limit}
\vskip 0.3cm
We now consider the light quark of a HL meson moves circularly around the heavy quark which is at rest in the center of the circular motion. So this will reduce the $p_r \rightarrow 0$ and $ v_\bot \rightarrow 1$. This reduction yields
\begin{eqnarray} 
S(v_\bot)=\frac \pi2
\label{eq.8}\\
f(v_\bot)=\frac \pi8
 \label{eq.9}
\end{eqnarray}

When pluging into the general HL RFT equations, we obtain

\begin{eqnarray} 
\frac Lr=\frac{\pi a r}{4}
 \label{eq.10}\\ 
E=\frac{\pi a r}{2}
\label{eq.11} 
\end{eqnarray}

Therefore we find the $`Nambu'$ slope for the HL meson is 

\begin{equation} 
{\alpha}^{'}_{HL} = \frac {L}{E^2}=\frac {1}{\pi a}
\end{equation} 

\vskip 0.3cm 

\subsection{Deep radial limit} 

\vskip 0.3cm

That the QCD string reduces to a time component vector potential was shown under the deep radial limit in which $L\ll E^2$. It is reasonable to apply the semi-classical quantization condition for a spherically symmetric system to quantize the QCD string,

\begin{equation} 
(n+ \frac 12)\pi = \int_{r_-}^{r_+} dr p_r = I
\label{aa}
\end{equation} 

where $r_+, r_-$ represent the radial distances at the turning points of the motion. Using this semi-classical quantization condition we obtain the following equation satisfied by the semi-classical TCV integral for HL meson,

\begin{equation} 
I =\frac{2 (e^2-L)\sqrt{(e^2+L)}}{e} J =(n+\frac12)\pi
\label{bb}
\end{equation} 

where 

\begin{equation} 
 J = \int_{0}^{1} dy \frac {\sqrt{(1-y^2)(1-m y^2)}}{1-h y^2} 
\label{jij}
\end{equation} 

We have defined that

\begin{equation} 
m = \frac{ e^2-L}{e^2+L} \hspace{ 0.5cm}  h = \frac{e^2-L}{e^2} \hspace{ 0.5cm}e^2=\frac{E^2}{4 a}
\end{equation} 

When evaluating Eq.(~\ref{bb}) we need replace the classical angular momentum with Langer correction to take into account the centrifugal singularity, namely $L \to l+\frac12$.

We give an exact result of this TCV integral for the heavy-light mesons by means of numerical method. The relation(\ref{bb}) of angular momentum and energy in ground state of the system when $k=0$ is shown in FIG.2.

\begin{figure}[ht]
\includegraphics[width=\columnwidth,angle=0]{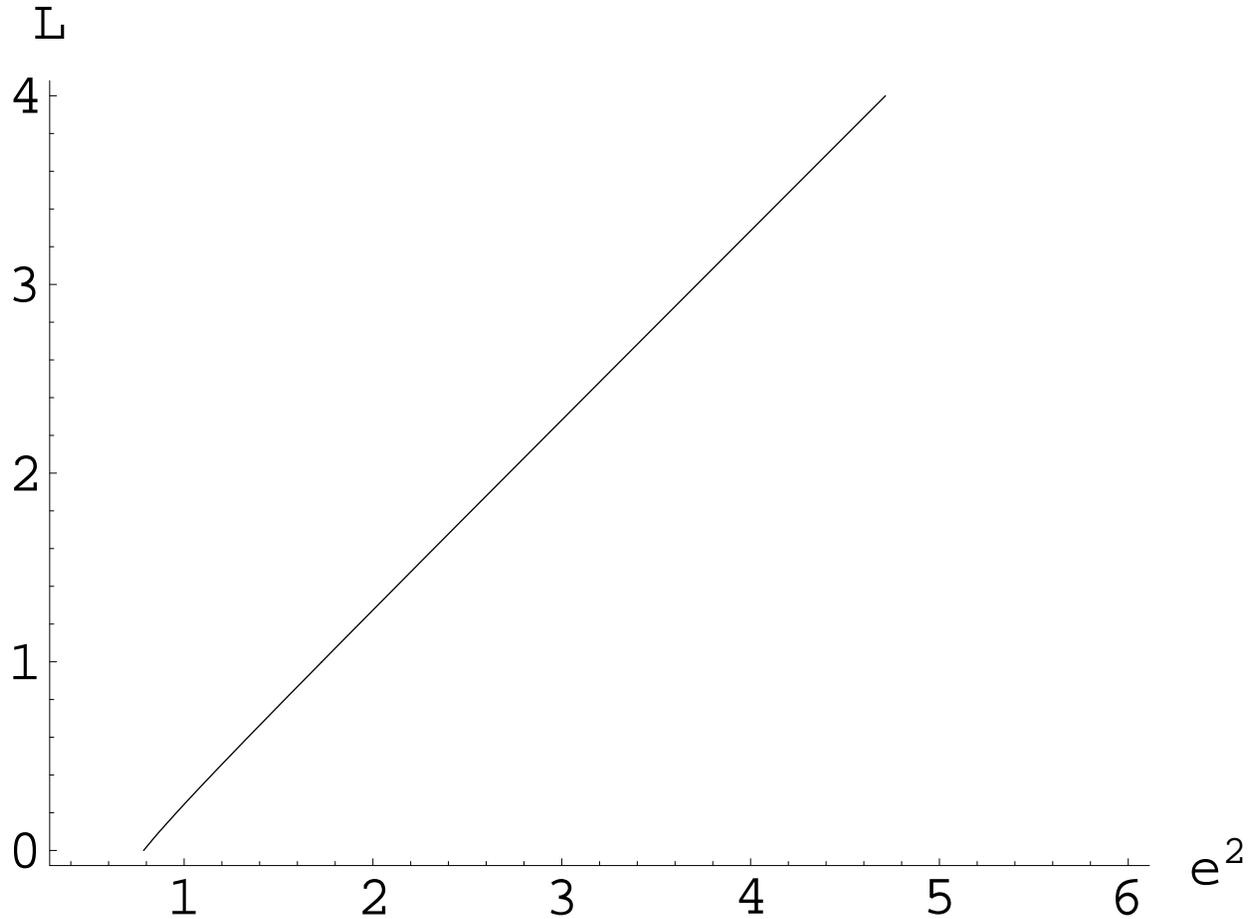}
\caption{Dependence of $L$ on $e^2$ in the ground state of a heavy-light meson. \label{fig:2}}
\end{figure}

To comprehend more about the variations of angular momentum with energy in different state of the quark system, we draw three states in FIG.3 corrresponding to $n = 0, 1$ and 2 respectively,

\begin{figure}[ht]
\includegraphics[width=\columnwidth,angle=0]{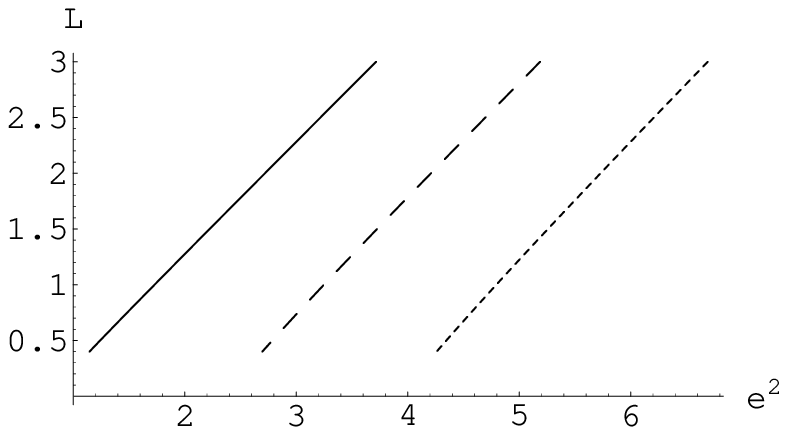}
\caption{The middle dashed line is for the first excited state, the right one is for the 2nd excited state while the solid line is for the ground state. \label{fig:3}}
\end{figure}

Now we are ready to turn on the color Coulomb potential $V(r)=-\frac{k}{r}$ to explored what effect the short range Coulomb potential has on our graphs. We find the new relation of angular momentum and energy in the system is

\begin{eqnarray}   
(n+ \frac 12)\pi&=\frac{\displaystyle 2 (e^2-L+k) \sqrt{(e^2+L+k)}}{\displaystyle e} \\
& \int_{0}^{1} dy \frac {\displaystyle \sqrt{(1-y^2)(1-\frac{ \displaystyle e^2-L+k}{\displaystyle e^2+L+k} y^2)}} {\displaystyle 1-\frac{e^2-L+k}{\displaystyle e^2} y^2} \nonumber 
\label{eq:a}
\end{eqnarray} 

According to the above equation, we may draw the graphs representing the relation of angular momentum and energy numerically. Considering the ground state of the quark system and choosing $k = 0, 0.2, 0.4$ separately, the modified curves due to the short range color Coulomb potential are shown in FIG.4. Note that for a fixed $L$, higher $k$ value curve gives the lower energy. This is due to the color Coulomb potential we take.

\begin{figure}[ht]
\includegraphics[width=\columnwidth,angle=0]{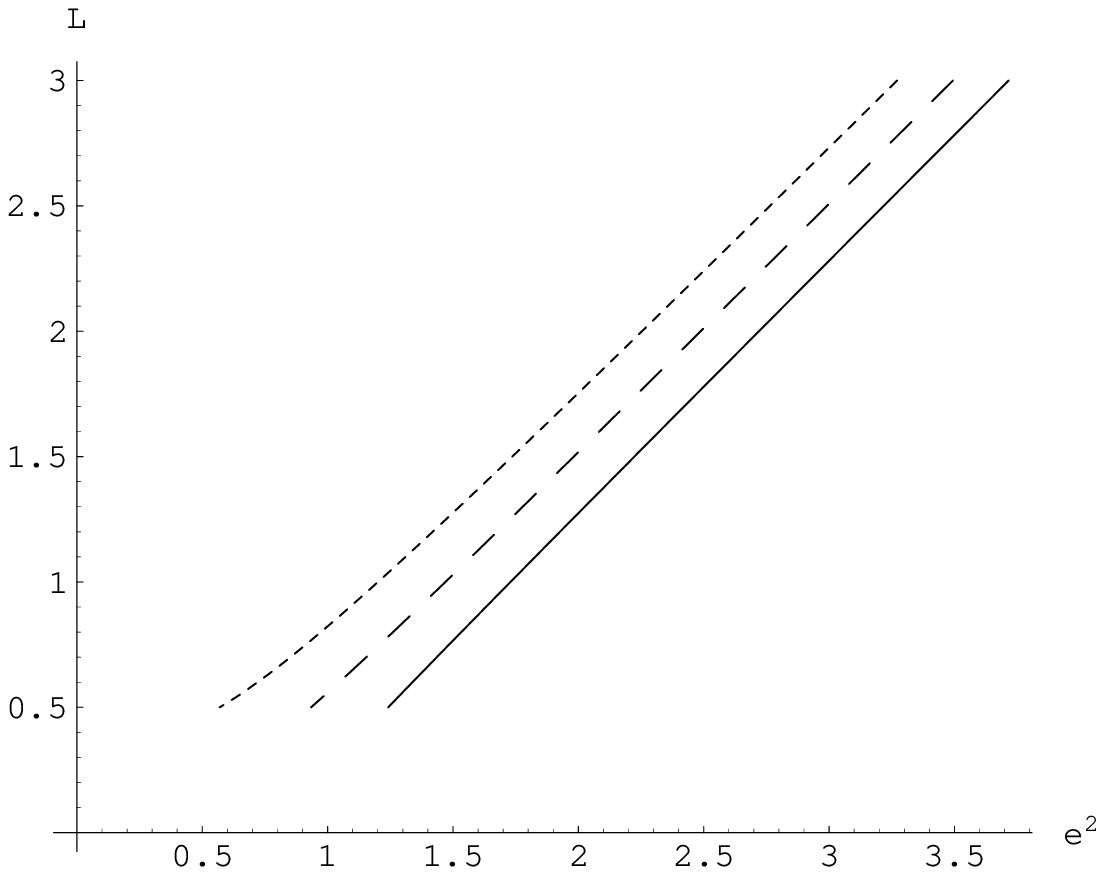}
\caption{The solid curve represents color Coulomb constant $k=0$, the left dashed curve corresponds to the $k=0.4$ while the middle dashed one is for $k=0.2$. \label{fig:4}}
\end{figure}

\section{Comparison of an analytic approximation solution with the exact numerical solution}

\subsection{An analytic approximation solution} 

We now derive an analytic approximation of RFT considering $L\ll E^2$ and $k=0$. To this purpose, we define 

\begin{equation} 
\beta = \frac {L}{e^2}
\end{equation} 

Since $\beta \ll 1$ in the deep radial limit, m and h may be represented with $\beta$ as 

\begin{eqnarray} 
m =\frac{1- \beta}{1+ \beta} \approx 1-2 \beta
 \label{eq.18}\\ 
h \approx 1- \beta
\label{eq.19} 
\end{eqnarray}

Plugging into Eq.(\ref{jij}) we obtain

\begin{equation} 
 J = \int^1_0 dy \frac {\displaystyle\sqrt{(1-y^2)(1-y^2+2 \beta y^2)}}{\displaystyle 1-y^2+\beta y^2} 
\end{equation}

Assuming that,
\begin{equation} 
1-y^2 = \frac{1}{1+x}
\end{equation}

then $J$ may be represented with the new parameters $x$ and $\beta$ as

\begin{equation} 
 J =\frac 12 \int^\infty_0 dx \frac{\displaystyle \sqrt{\frac{\displaystyle (1+2 \beta x)}{\displaystyle x (1+x)^3}}}{\displaystyle 1+ \beta x} 
\end{equation}

Next expand $J$ with taylor expansion and drop terms of second order in $\beta$ or higher, so 

\begin{equation} 
 J= J(0)+ \frac {\partial J}{\partial \beta} \beta + \bigcirc (\beta^2)
\end{equation}
 
For $\beta=0$,

\begin{equation} 
 J(0) = \frac 12 \int^\infty_0 dx \sqrt{\frac{\displaystyle 1}{\displaystyle x(1+x)^3}} = 1
\end{equation}

and 

\begin{eqnarray}
\left.\frac {\displaystyle \partial J}{\displaystyle \partial \beta}\right|_{\beta=0} & = & \frac 12 \int^\infty_0 dx \sqrt{\frac{\displaystyle 1}{\displaystyle x(1+x)^3}} \frac{\displaystyle \partial (\frac {\displaystyle \sqrt{1+2 \beta x}}{\displaystyle 1+ \beta x})}{\displaystyle \partial \beta},  \nonumber  \\  
& = & \frac 12 \int^\infty_0 dz\frac{\displaystyle 1}{\displaystyle \sqrt{(1+2 z)} (1+z)^2} = \frac 12-\frac {\pi}{4}
\label{eq:aa}
\end{eqnarray}

where we let $z=\beta x$. Therefore when  $\beta \ll 1$, we find to first order in $ \beta$

\begin{eqnarray}
I & = & \frac{\displaystyle  2(e^2-L) \sqrt{e^2+L}}{\displaystyle e} J= 2 e^2 (1-\beta)(1+\beta)^\frac 12 [1+(\frac 12 -\frac {\pi}{4})\beta\,] \nonumber  \\  
& = & 2 e^2 (1-\frac {\pi}{4}\beta)
\label{eq:ab}
\end{eqnarray}

After substituting $e^2$ with $\frac{E^2}{4a}$, we obtain an analytical approximation of the heavy-light semi-classical time component vector potential integral,

\begin{equation} 
I =\frac {E^2}{2a} (1-\frac {\pi}{4}\beta)
\end{equation} 

Applying the semi-classical quantization condition for a spherically symmetric system  and replacing the classical angular momentum with Langer correction to take into account the centrifugal singularity, we immediately obtain the spectroscopic relation

\begin{equation} 
L + 2 n+ \frac 32 =\frac{E^2}{\pi a} 
\label{cc}
\end{equation}

which is an excellent approximation to its numerical solution shown in section \ref{consis}.

\subsection{Consistency of the analytic approximation solution and the exact numerical result} \label{consis}

We are now ready to compare this analytic approximation result with the exact numerical one. We draw the six figures(from FIG.5 to FIG.10) for every other $n$ starting from 0 to 10. From these figures we easily identify that the approximations are more accurate along with the increasement of n since the deep radial limit is much more satisfied. 

\begin{figure}[Ht]
\includegraphics[width=\columnwidth,angle=0]{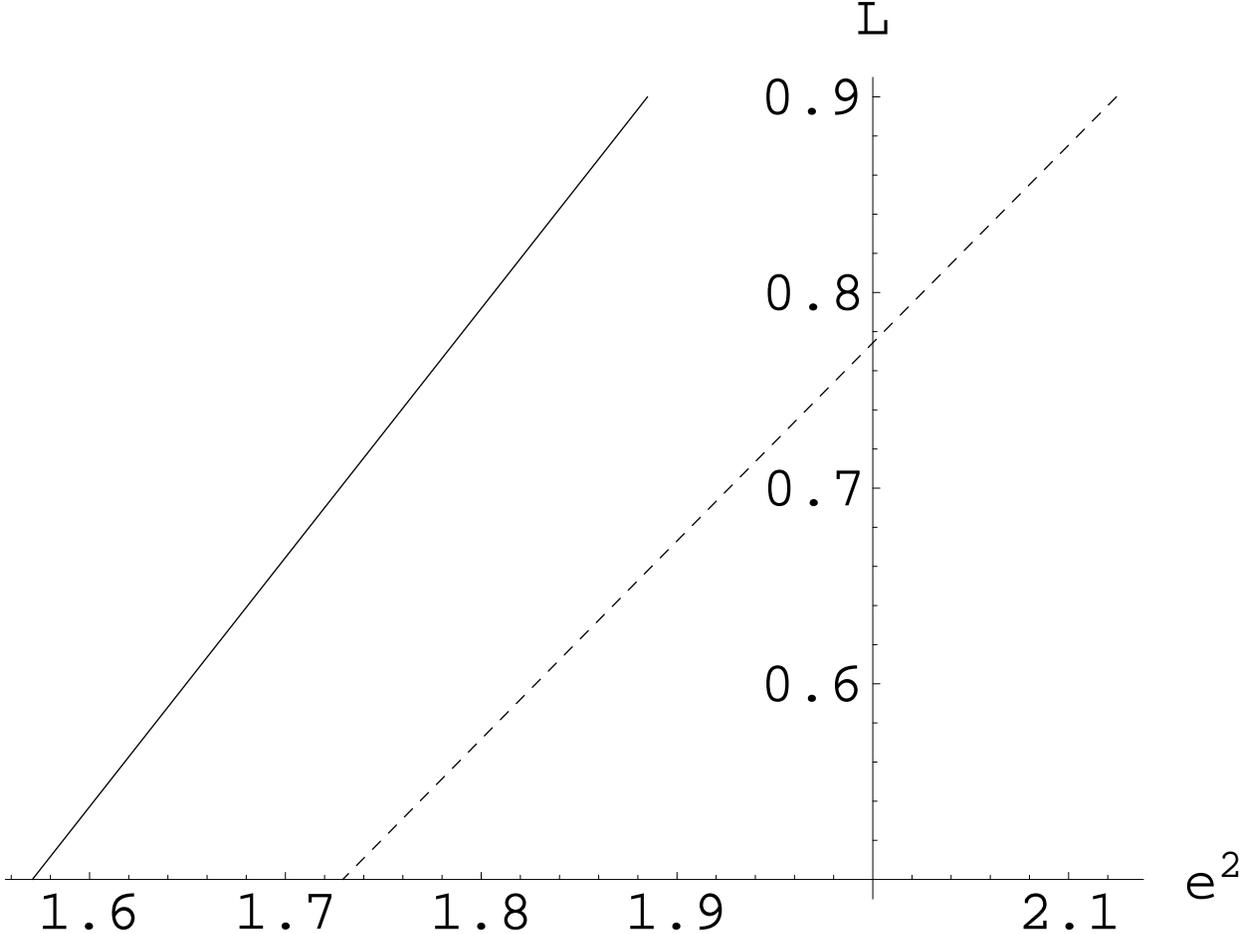}
\caption{The ground state n=0 \label{fig:5}}
\end{figure}

\begin{figure}[Ht]
\includegraphics[width=\columnwidth,angle=0]{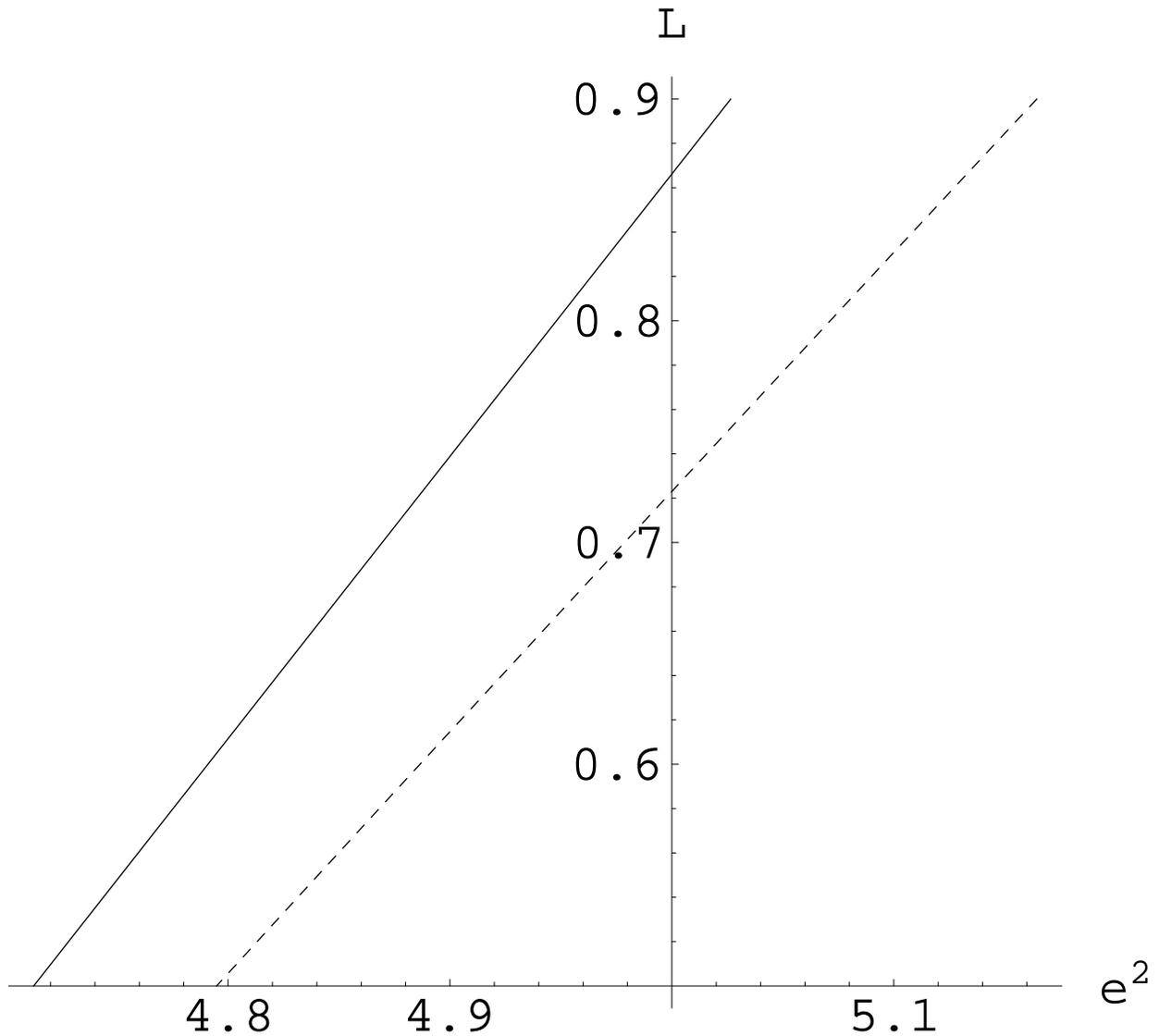}
\caption{The 2nd excited state \label{fig:6}}
\end{figure}

\begin{figure}[Ht]
\includegraphics[width=\columnwidth,angle=0]{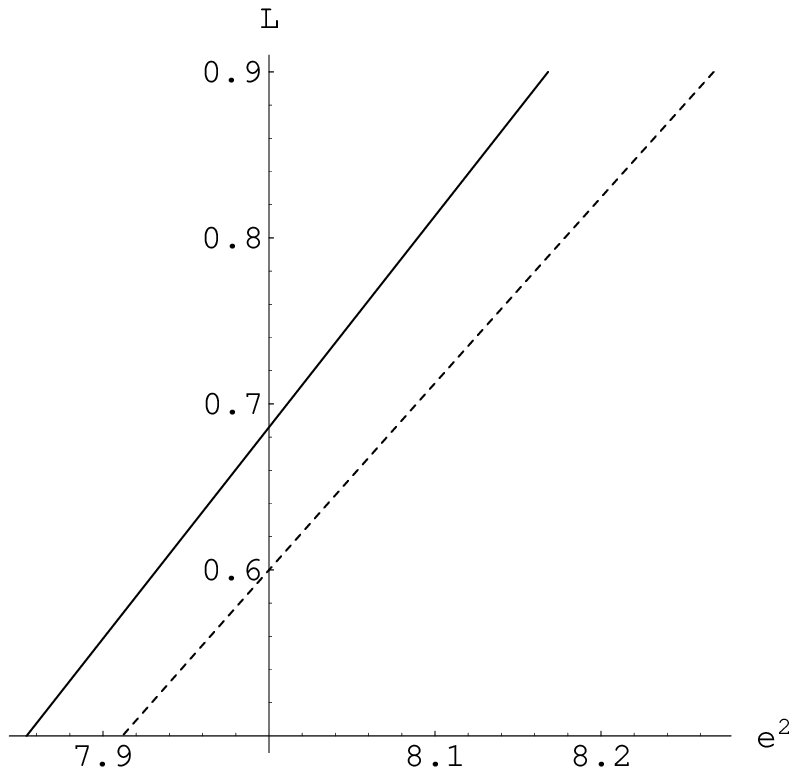}
\caption{The 4th excited state \label{fig:7}}
\end{figure}

\begin{figure}[Ht]
\includegraphics[width=\columnwidth,angle=0]{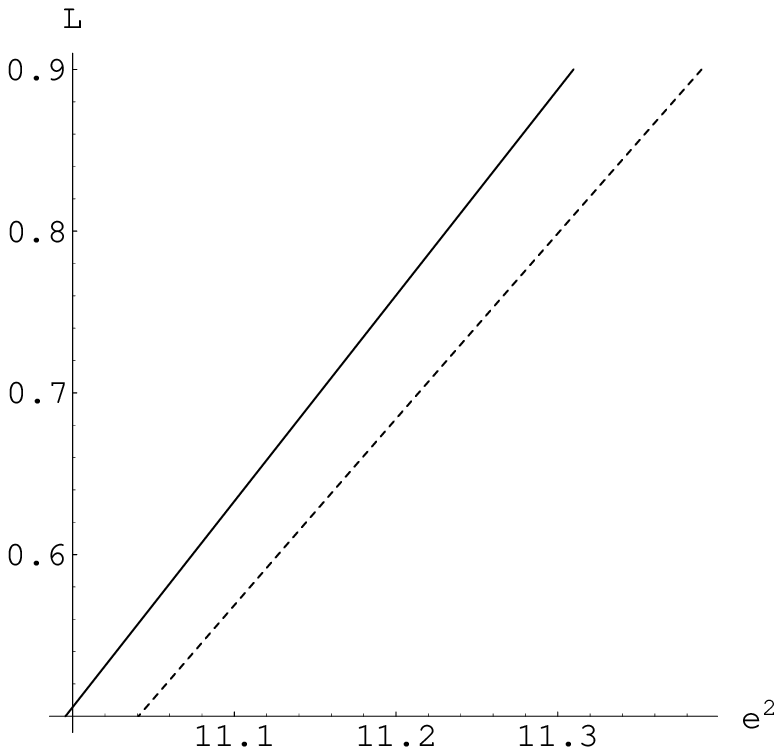}
\caption{The 6th excited state \label{fig:8}}
\end{figure}

\begin{figure}[Ht]
\includegraphics[width=\columnwidth,angle=0]{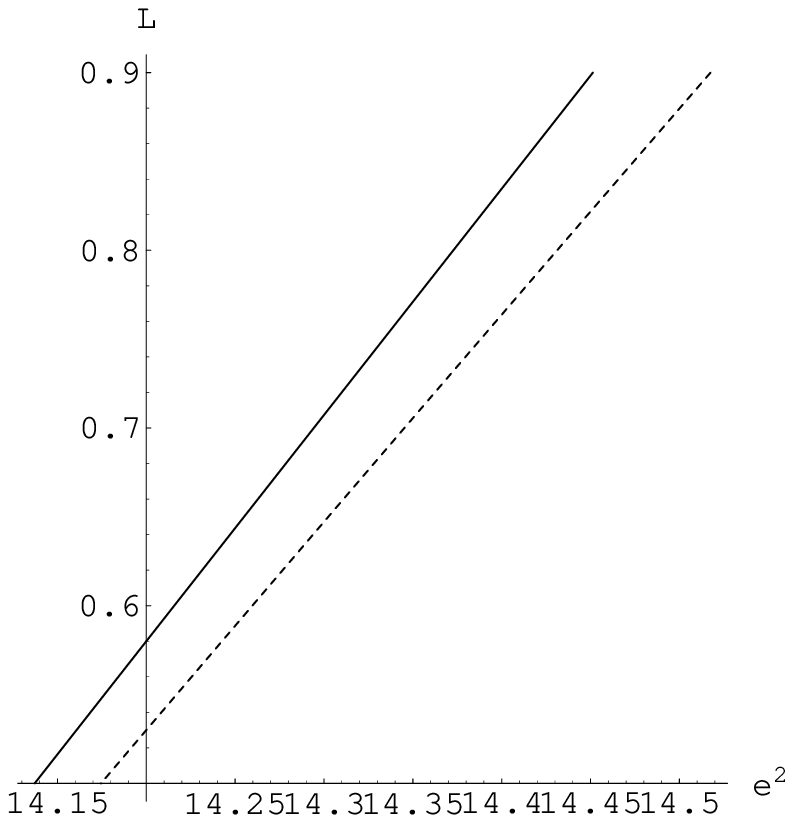}
\caption{The 8th excited state \label{fig:9}}
\end{figure}

\begin{figure}[Ht]
\includegraphics[width=\columnwidth,angle=0]{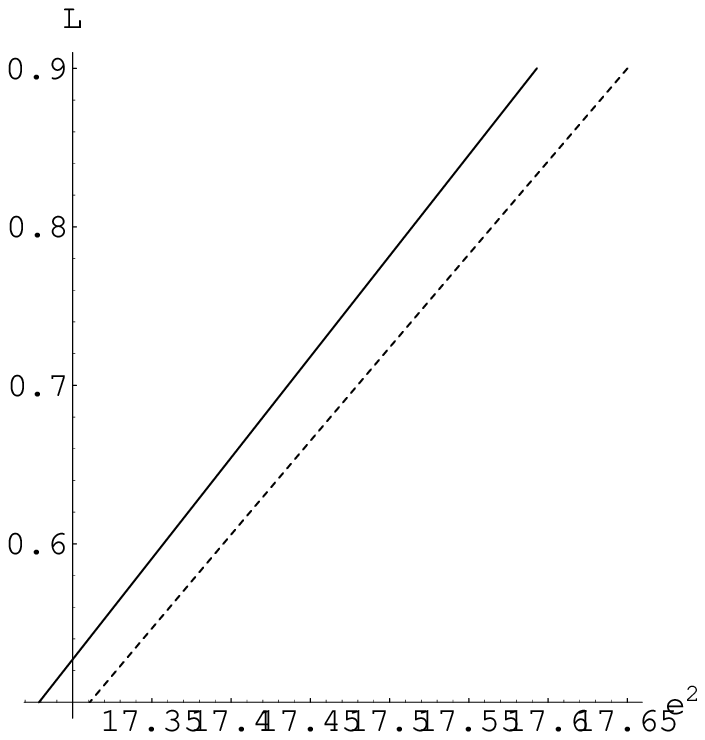}
\caption{The 10th excited state \label{fig:10}}
\end{figure}

\section{Summary and Conclusion}

We have studied the scalar potential, time component vector potential and flux tube quark confinements. The Lorentz scalar quark confinement has a long history and is still widely used despite its well-known theoretical faults.  We point out here that the predictions of scalar confinement also conflict directly with experiment.  We investigate the dependence of heavy-light meson mass differences on the mass of the light quark.  In particular, we examine the strange and non-strange $D$ mesons. Under two limit we also solve the relativistic flux tube equations. We have presented the exact numerical solution for HL mesons and compared the numerical solution with an analytical approximation solution.

The conclusions are that spin splittings are independent of the light quark mass value.  We then can extract the change (\ref{Diff}) in the spin-averaged 1\textit{P}$-$1\textit{S} level difference as one replaces the non-strange by a strange light quark. We compare the predictions of the scalar potential, time component vector potential and flux-tube quark confinement scenarios with experimental results. We conclude that flux-tube confinement works well while both scalar and time component vector confinement fail badly to explain the experimental data. We observe therefore that scalar confinement has at least one specific point of disagreement with experiment. This complements the theoretical disagreements with QCD mentioned in the introduction. For the relativistic flux tube model, from the comparison of exact numerical solution with the analytic approximation solution for heavy-light mesons, we see that they are more in agreement with each other for higher excited states since the deep radial limit is better satisfied. The relativistic flux tube model gives the correct universal relations for HL meson spectroscopies.

\section*{Acknowledgment}

The author would like to thank M. Olsson, C. Goebel and S. Veseli for their helpful discussions and collaborations. This work was supported in part by the U.S. Department of Energy under Contract No.~DE-FG02-95ER40896.     

\section*{Appendix    Derivation of the expectation values}

We now derive the expectation value of $r^p$ with
the harmonic oscillator wavefunctions (\ref{soln0}),
\begin{equation}
\langle r^p \rangle =N_{n,\ell}^2\> 
\int_0^{\infty} dr\,r^{2\ell+2+p}\, e^{-ar^2}\,
\left[L_{n-1}^{\ell+\frac{1}{2}}(ar^2)\right]^2 .   \label{exr}
\end{equation}
A change of integration variable to the dimensionless combination
$z=a\,r^2$ yields 
\begin{equation}
2\,a^{\ell +\frac{3}{2}+\frac{p}{2}}\, \langle r^p \rangle = N_{n,\ell}^2\,
\int_0^\infty dz\,z^{\ell
+\frac{1}{2}+\frac{p}{2}}\,e^{-z}\,\left[L_{n-1}^{\ell
+\frac{1}{2}}(z)\right]^2 .        \label{exz}
\end{equation}
It is helpful to use the Chu-Vandermonde sum formula \cite{GR},
\begin{equation}
L_{n-1}^\alpha (z)=\sum_{j=1}^n ~\frac{(\alpha -\beta)_{n-j}}{(n-j)!}
~L_{j-1}^\beta (z) ,
\label{CV}
\end{equation}
where the Pochhammer symbol $(z)_N$ is defined as,
\begin{eqnarray}
(z)_N =z(z+1)\cdots (z+N-1) & = & \frac{\Gamma(z+N)}{\Gamma(z)},\label{poch}\\
(z)_0 & = & 1 .
\end{eqnarray}

With the choices
\begin{eqnarray}
             \alpha & = & \ell +\frac{1}{2} , \\  
              \beta & = & \ell +\frac{1}{2} + \frac{p}{2} ,    \\  \label{def1}
      \alpha -\beta & = & - \,\frac{p}{2} ,
\end{eqnarray}
we substitute Eq.~(\ref{CV}) into (\ref{exz}) and use the orthonormality
relation for Laguerre polynomials \cite{GR}
\begin{equation}
\int_0^\infty dz\,z^\beta \, e^{-z} \, L_j^\beta (z) \, L_{j'}^\beta (z) =
\frac{\Gamma(j+\beta + 1)}{j!} \, \delta_{jj'}   ,   \label{orthog} 
\end{equation} 
to obtain our general result,
\begin{equation}
\langle r^p \rangle = a^{-\frac{p}{2}} \, \sum_{j=1}^n
\,\left[\left(-\frac{p}{2}\right)_{n-j}\right]^2\, \pmatrix{n-1 \cr j-1}
\frac{\Gamma (j+\ell+\frac{1}{2}+\frac{p}{2})}{\Gamma
(n+\ell+\frac{1}{2})(n-j)!} .
\label{finalexpectation}
\end{equation}
In Eq.~(\ref{finalexpectation}) we use the notation for the binomial coefficients,
\begin{equation}
\pmatrix{n\cr m} = \frac{n!}{m!(n-m)!} .      \label{binomial}
\end{equation}
The specific result that is required in Section 2 is for the
ground state 
($n=1$) is
\begin{equation}\label{n=1}
\langle r^p \rangle\strut_{n=1} = a^{-\frac{p}{2}} ~\frac{\Gamma (\ell
+\frac{3}{2}+\frac{p}{2}) }{\Gamma(\ell+\frac{3}{2})} 
\end{equation} 
Some cases of direct interest are
\begin{eqnarray}
  & p = -1: \qquad &\left\langle {r^{-1}} \right\rangle = \sqrt{a}\,
              \frac{\Gamma(\ell+1)}{\Gamma(\ell+\frac{3}{2})} , \\ 
  & p =  0: \qquad &\langle 1 \rangle = 1 , \\
  & p =  1: \qquad &\langle r \rangle = \frac{1}{\sqrt{a}}\,
              \frac{\Gamma(\ell+2)}{\Gamma(\ell+\frac{3}{2})} , \\ 
  & p =  2: \qquad &\langle r^2 \rangle = \frac1a \left(\ell +\frac{3}{2}\right) .
\end{eqnarray}

\end{document}